\documentclass[prx,aps,twocolumn,superscriptaddress,notitlepage]{revtex4-2}
\usepackage{amssymb,amsfonts,amsmath,amstext,graphicx,hyperref}
\usepackage{tikz}
\usepackage{float}
\usepackage{graphicx}
\usepackage[normalem]{ulem}
\hypersetup{
	colorlinks=true,
	linkcolor=blue,
	filecolor=magenta,      
	urlcolor=cyan,
}
\usepackage{xcolor}

\usepackage[normalem]{ulem}
\newcommand{\cut}[1]{}

\usepackage{physics}

\begin{document}

\title{Non-Ideal Measurement Heat Engines}

\author{Abhisek Panda}
\affiliation{Department of Physics, Indian Institute of Technology Bombay, Powai, Mumbai 400076, India.}
\author{Felix C. Binder}
\affiliation{School of Physics, Trinity College Dublin, Dublin 2, Ireland.}
\affiliation{Trinity Quantum Alliance, Unit 16, Trinity Technology and Enterprise Centre, Pearse Street, Dublin 2, D02 YN67, Ireland.}
\author{Sai Vinjanampathy}
\email{sai@phy.iitb.ac.in}
\affiliation{Department of Physics, Indian Institute of Technology Bombay, Powai, Mumbai 400076, India.}
\affiliation{Centre for Quantum Technologies, National University of Singapore, 3 Science Drive 2, 117543 Singapore, Singapore.}
\affiliation{Centre of Excellence in Quantum Information, Computation, Science, and Technology, Indian Institute of Technology Bombay,
Powai, Mumbai 400076, India.}

\date{\today}

\begin{abstract}
We discuss the role of non-ideal measurements within the context of measurement engines by contrasting examples of measurement engines which have the same work output but with varying amounts of entanglement. Accounting for the cost of resetting, correlating the engine to a pointer state and also the cost of cooling the pointer state, we show that for a given work output, thermally correlated engines can outperform corresponding entanglement engines. We also show that the optimal efficiency of the thermally correlated measurement engine is achieved with a higher temperature pointer than the pointer temperature of the optimal entanglement engine. 
\end{abstract} 

\maketitle

\section{Introduction}\label{section}
Quantum heat engines are composed of a system coupled to one or more environments such that at least one party exploits quantum energetic coherence. Quantum engines with both thermal \cite{alicki1979quantum,quan2007quantum} and non-thermal  \cite{manzano2016entropy,niedenzu2018universal,agarwalla2017quantum} reservoirs, energetic coherence \cite{hammam2021optimizing,klatzow2019experimental,uzdin2016coherence,scully2010quantum,binder2015quantum,kammerlander2016coherence} and entanglement \cite{huang2013special,PhysRevA.101.012315,PhysRevB.101.081408} have been studied to understand how quantum engines differ from their classical analogues. In this context, it is important to consider carefully the role of quantum measurement, given its central role in quantum theory. Alongside a resource-theoretic characterization of thermodynamic quantum operations \cite{Janzing2000,brandao2013resource,faist2015gibbs,Brandaoa2015,lostaglio2016resource,Muller2018}, a careful accounting of the energetic constraints on quantum measurements is needed to understand their role in non-equilibrium engines.
Just as in the case of thermal baths, measurements can result in an apparent change of both entropy and energy with the working fluid. Such measurements and feedback have been discussed in the context of thermal engines \cite{seah2020maxwell,kim2011quantum}. Measurement and feedback have been used in the classical and quantum context to elucidate the role of information and correlations in the second law of thermodynamics \cite{kieu2006quantum,PhysRevLett.118.260603,vinjanampathy2016correlations,francica2017daemonic,santos2019role,jacobs2012quantum,gherardini2020stabilizing}. While engines have been shown to exploit both thermal and non-classical correlations, measurement models have not considered the thermodynamic infeasibility of projective measurements due to the inconsistency of the third law of thermodynamics and preparation of pure state pointers \cite{wilming2017third,scharlau2018quantum,kieu2019principle}. It was shown recently that when thermodynamic feasibility is taken into account, ideal measurements which correlate the system states to the pointer states become impractical and two varieties of ``non-ideal” measurements emerge as feasible measurement models \cite{guryanova2020ideal,debarba2019work}.

This hence raises the natural question as to whether non-ideal measurements and the enhanced performance of quantum thermal machines whose operation includes a measurement stage are compatible. Here, we consider two models of quantum engines and show that indeed non-ideal measurements can still be used to construct engines which consume either thermal or non-classical correlations as fuel. Furthermore, we show that we can always construct a thermal engine that outperforms the entanglement engine presented below in terms of efficiency for the same value of average output work.
We begin the next section with a brief review of ideal and non-ideal measurements. In Sec.~\ref{sec:III} we describe the working of a generalized measurement engine. The findings of Sec.~\ref{sec:III} are exemplified in Sec.~\ref{sec:IV} where we consider a two-qubit measurement engine and compare the performance with entangled and classically correlated initial states. Finally, we discuss our results in Sec.~\ref{sec:V}.

\section{Non-Ideal Measurements and Work Cost}
It is well known that ideal projective measurements on quantum systems using finite resources are incompatible with thermodynamics \cite{guryanova2020ideal,debarba2019work}. Hence, it is necessary to take the resource cost of measurement into account and consider a description of measurements that is physical. This inclusion of thermodynamics gives rise to the notion of ``non-ideal" measurements, which we briefly review. 

Consider a generic quantum system $\rho_{S}$ and a measuring device (pointer) $\rho_{P}$. A measurement of the system is achieved by coupling it to the pointer and allowing the joint quantum state to undergo a correlating evolution from $\rho_{S} \otimes \rho_{P}$ to a correlated state $\rho_{SP}$ \cite{sewell2007can}. Three fundamental features define an ideal projective measurement $\mathbb{I} \otimes \Pi_{i}$, which infers the system state from the measured pointer state \cite{guryanova2020ideal}. The first property desired for ideal measurements is \textit{unbiasedness}, defined as the pre-measurement probability statistics of the system being accurately reflected by the post-measurement pointer statistics, namely

\begin{equation} 
\Tr[\mathbb{I} \otimes \Pi_{i} \rho_{SP}] = \rho_{ii}, \ \ \forall \ i,
\label{unbiased}
\end{equation}
where $\rho_{ii} = \Tr[\ketbra{i}_S \rho_S]$ are the diagonal elements of $\rho_S$.

Secondly, the measurement should be \textit{faithful}, defined by the property that the post-measurement system state should correspond to the outcome of the pointer, namely
\begin{equation} 
\sum_{i} \Tr[\ketbra{i} \otimes \Pi_{i} \rho_{SP}] = 1 \ \  .  
\label{faithful} 
\end{equation}
Thirdly, ideal measurement should be \textit{non-invasive}, defined by the property that the measurement interaction should not change the measurement statistics for the system, i.e., 
\begin{equation} 
\Tr[\ketbra{i}\otimes \mathbb{I} \rho_{SP}] = \rho_{ii}, \ \ \forall \ i. 
\label{non-invasive}
\end{equation}
\begin{figure}[]
\includegraphics[width=\columnwidth,height=0.65\columnwidth]{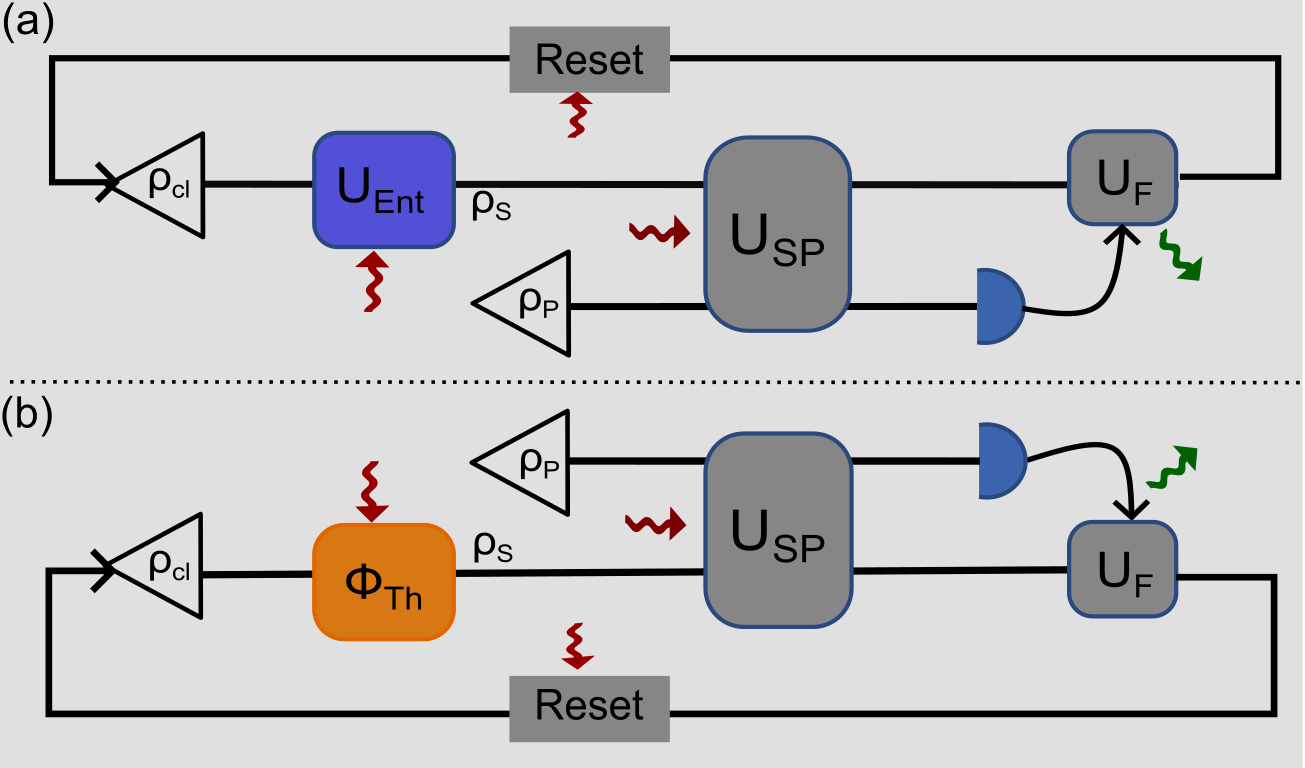}
\caption{The measurement engine cycle with (a) entangled state (EME) and (b) thermally correlated state (TCME) includes the creation of system state~$\rho_S$, correlating it with the pointer system in state~$\rho_P$ (via $U_{SP}$), measurement of the pointer followed by feedback and reset of the system to state~$\rho_{cl}$. The energy-consuming steps are indicated by red arrows and work extraction is indicated by green arrows.}
\label{fig:2}
\end{figure}

\noindent The measurement can be faithful only if the rank of the pointer obeys $\mathrm{rank(\rho_{P})} \leq n_p/n_s$, where $n_p$ and $n_s$ are dimension of pointer and system respectively (both assumed to be finite wiithout loss of generality)\cite{guryanova2020ideal}. Thus an ideal measurement requires a non-full rank pointer state, but according to third law of thermodynamics, preparation of such states require infinite resources~\cite{taranto2021landauer}, which makes all practical measurements non-ideal. This leads to the conclusion that a practical measurement on a thermodynamic system could be unbiased or non-invasive but not both, as both properties together would also imply faithfulness\cite{guryanova2020ideal}. In either of these cases, after imposing the condition, we can quantify the faithfulness of the measurement by following Eq.\eqref{faithful}. This can be done by defining the correlation function $C_{max}(\rho_{SP}) = \sum_{i} \Tr[\ketbra{i} \otimes \Pi_{i} \rho_{SP}] < 1$ which quantifies the probability of joint measurement of the system and pointer in the same state.

In summary, both ideal measurements and pure input states are thermodynamically incompatible since they require infinite resources \cite{masanes2017general,wilming2017third,clivaz2019unifying,allahverdyan2011thermodynamic, vinjanampathy2016quantum,taranto2021landauer}. Thus, quantum correlations produced either as a consequence of pure initial states or non-full rank projective measurements do not sufficiently inform us of the role of real measurements in quantum thermal machines. We address this by considering two different types of measurement engines, an entanglement-based measurement engine (EME) using entangled states as initial states and a thermally correlated measurement engine (TCME) using initial states with classical correlations compliant with a thermal distribution, both constructed from full rank states and operated by non-ideal measurements as shown in Fig.~\ref{fig:2}. In the next section we will look at full-rank quantum engines driven by unbiased and maximally faithful measurements. Likewise, we also inspect the role of non-invasive and maximally faithful measurements in quantum engines.


\section{\label{sec:III}Bipartite  Measurement Engines}
\noindent To investigate the nature of quantum correlations on the performance of quantum thermal machines, we consider a generic bipartite quantum measurement engine which describes both EMEs and TCMEs. Let the working medium consist of two $n_s$ level systems, A and B, with the joint Hamiltonian $H=H_{loc}+H_{int}$. The term $H_{loc}$ is the local Hamiltonian of both qudits defined as 
\begin{equation}
    H_{loc} = \sum_{i=0}^{n_s-1} E^{A}_i (\ketbra{i}{i} \otimes I) + E^{B}_i (I \otimes \ketbra{i}{i}),\label{hloc}
\end{equation}
where $\{E_i^A\}$ and $\{E_i^B\}$ are energy eigenvalues of subsystem A and B respectively.
As discussed below, appropriate feedback will extract the difference of energy eigenvalues as work, hence without loss of generality we can choose 
\begin{equation}E^B_{i}-E^B_{j} > E^A_{i}-E^A_{j}\geq 0 \ \ \forall \ i>j.
\end{equation}
The interaction term $H_{int}$, generates entanglement between the two qudits. It is generically defined as
\begin{equation}
    H_{int}=  \sum_{i=0}^{n_s-1}\sum_{j=0}^{n_s-1}  \dfrac{g_{ij}}{2} (\ketbra{ij}{ji} + \ketbra{ji}{ij}),
    \label{hint}
\end{equation}

\noindent where $g_{ij}$ is the coupling strength. This interaction can be switched on to prepare initial states in the EME. There is a cost associated for switching $H_{int}$ that needs to be taken into account while running the EME. Both measurement engines operate in four stages which we briefly review~\cite{mohammady2017quantum,ito2019generalized,jordan2020quantum,PhysRevE.98.042122,PhysRevE.96.022108,PhysRevLett.120.260601,PhysRevLett.118.260603,PRXQuantum.2.040328,thingna2020quantum}.\\


\noindent\textbf{(i) State preparation:}  The initial states for the measurement engine are prepared from a reference product state~$\rho_{cl}$ (assumed to be diagonal). In the EME case, the entangled states are prepared by evolving the reference state with the interaction Hamiltonian, which is then turned off after creating the desired entangled state $\rho_S$. For TCME, on the other hand, classical correlations are induced by a thermal CPTP map. The  energy cost associated with the preparation in both engines is given by $E_{prep} = \Tr [H_{loc}(\rho_{S}-\rho_{cl}) ].$ \\ 

\noindent \textbf{(ii) Pointer correlation:} A quantum system is measured with the help of a pointer degree of freedom ~\cite{von2018mathematical}. Since we are measuring a bipartite system, for simplicity, we consider the pointer system to be bipartite as well, governed by Hamiltonian 
\begin{equation}H_P=\sum_{i=0}^{n_p-1} E^{PA}_i (\ketbra{i}{i} \otimes I) + E^{PB}_i (I \otimes \ketbra{i}{i}).
\end{equation} 
Here $E_i^{PA}$ and $E_i^{PB}$ are the energy gaps of the pointer subsystems measuring systems \textbf{A} and \textbf{B}, respectively and without loss of generality $n_p$ is assumed to be an integer multiple of $n_s$. For thermodynamic consistency, we consider a full rank pointer $\tau(\beta)=e^{-\beta H_P}/\text{Tr}(e^{-\beta H_P})$, at inverse temperature $\beta$. In order to increase measurement accuracy, the pointer is initially cooled from $\beta\rightarrow\beta'>\beta$. This process consumes free energy difference~$ (E_{cool})$~\cite{clivaz2019unifying} which is same for both TCME and  EME. The pointer is then correlated with the system, and the nature of the correlation matrix determines if the measurement is unbiased or non-invasive \cite{guryanova2020ideal}. The cost associated with correlating the system to the pointer is given by
\begin{equation}
    \begin{aligned}
   \hspace{-0.2cm} E_{corr}&=Tr[(H_{loc}+H_{P})\{\rho_{SP}-\rho_S\otimes\tau_P\}].
   \end{aligned}
   \label{ecor}
   \end{equation}
   Here $\rho_{SP}=U_{corr}(\rho_{S}\otimes\tau_P) U^\dagger_{corr}$ and $U_{corr}$ is the correlation matrix. The choice of $U_{corr}$ is made such that the measurement is maximally faithful as described below.\\

   
\noindent \textbf{(iii) Work extraction:} The measurement of the primary system is performed through the pointer. Based on the outcome of this pointer measurement we then apply feedback to system. If the pointer measurement implies that subsystem \textbf{B} is in a higher energy state than subsystem \textbf{A}, we apply a swap operation and extract the work from the system. On the other hand, if the pointer reads subsystem \textbf{A} to be in higher energy state than subsystem \textbf{B}, no feedback is applied. After extracting work, the postselected state will be given by
\begin{equation}
    \begin{aligned}&\Psi(\rho_{SP})=\sum_{l=0}^{n_P-1}\sum_{k \geq l}^{n_P-1} I\otimes|kl\rangle\langle kl| \rho_{SP} I\otimes|kl\rangle\langle kl|\\&+\sum_{l=0}^{n_P-1}\sum_{k < l}^{n_P-1} U_{swap}\otimes|kl\rangle\langle kl|\rho_{SP} U^\dagger_{swap}\otimes|kl\rangle\langle kl|.
    \end{aligned}
\end{equation}
Here the $U_{swap}$ operator swaps the eigenstates of system \textbf{A} and \textbf{B}, i.e., $U_{swap}\ket{ij}=\ket{ji}, \forall i,j$.
    The extracted work amounts to 
    \begin{equation}
    \begin{aligned}
    &W=\text{Tr}\left[\left(H_{l o c}+H_{P}\right)\left(\Psi\left(\rho_{SP}\right)-\rho_{S P}\right)\right].
    \label{Ework}
    \end{aligned}
    \end{equation} 
The unavailability of non-full rank states translates to the fact that measurements are never perfectly faithful. This results in imperfect correlations where a single pointer state can point to various system states. The imperfect correlation between system and pointer it turn results in incorrect feedback to the system. Such incorrect feedback may correspond to either false positive or false negative reading of the pointer. A false positive outcome occurs when the pointer inaccurately reads that qudit \textbf{B} is in a higher excited state than qudit \textbf{A}. Thus applying a swap feedback results in energy consumption instead of work extraction. On the other hand, a false negative outcome occurs when qudit \textbf{B} is in a higher excited state than qudit\textbf{ A}. In this scenario, a swap operation can extract work from the system, but the pointer detects otherwise. This leads to lack of swap feedback and loss of work extraction. Thus the amount of work extracted (given by Eq.(\ref{Ework})) with non-ideal measurement will be less than that obtained with ideal measurement.\\

\noindent \textbf{(iv) Reset:} After extracting work we reset the pointer to the thermal state but as the free energy of such process is negative we consider the cost of resetting the pointer to be zero. To repeat the engine cycle again we reset the system to the reference state~$\rho_{cl}$. The minimum energy used to reset the system is given by 
    \begin{equation}
    \begin{aligned}
        &E_{reset}=\text{Tr}[ H_{loc} \left(\rho_{cl}-\text{Tr}_P\left[\Psi\left(\rho_{SP}\right)\right]\right)].
        \label{ereset}
    \end{aligned}
    \end{equation}

\bigskip
\noindent As the local Hamiltonian is diagonal in the measurement basis, the work extraction and the cost associated will depend on $P_{SP}(ijkl)$, the joint system-pointer probability of the system being in $\ket{ij}$ and the pointer reading $\ket{kl}$. A non-invasive measurement can be realized if the correlation operation is of the form $U_{corr}=\sum_{nm}|nm\rangle \langle nm|\otimes \Tilde{U}^{nm}$. For this case, we obtain 
\begin{equation}
P_{SP}(ijkl)=\left\langle i j\left|\rho_{S}\right| i j\right\rangle\langle kl|(\Tilde{U}^{ij}\tau_P\Tilde{U^\dagger}^{ij})|kl\rangle.
\end{equation}
Similarly, a measurement will be unbiased if $U_{corr}=\sum_{nmop}|nm\rangle \langle op|\otimes \left|op\right\rangle\left\langle nm\right|\Tilde{U}^{op}$, which will give 
\begin{equation}
P_{SP}(ijkl)=\left\langle kl\left|\rho_{S}\right| kl\right\rangle\langle ij|(\Tilde{U}^{kl}\tau_P\Tilde{U^\dagger}^{kl})|ij\rangle.
\end{equation}
The faithfulness of the measurement depends on the unitary matrix $\Tilde{U}^{ij}$. Maximally faithful measurement is achieved by using $\Tilde{U}^{ij}=\Tilde{U}^{i}\otimes\Tilde{U}^{j}$ where
\begin{align}
\Tilde{U}^{i}&=I-\sum_{x=0}^{\nu-1}\left(\ket{x}-\ket{\nu~i+x}\right)\left(\bra{x}-\bra{\nu~i+x}\right).
\end{align}
Here $\nu={n_p}/{n_s}$ is assumed to be an integer without loss of generality \cite{guryanova2020ideal}. For both types of measurement, we observe that the performance only depends on diagonal elements of the system density matrix. This indicates that while diagonal distributions preserve the work output, the impact of off-diagonal correlations on the engine performance is irrelevant. In the next section, we will explore the role of correlations on the performance of quantum thermal machines by studying two-qubit measurement engines powered either by quantum or thermal correlations, respectively.

\section{\label{sec:IV} Two qubit Measurement engine}
To compare and contrast varieties of correlations, we begin our study with an EME driven by full rank states. To do this, we modify a recently proposed \cite{bresque2021two,huang2020two} two qubit measurement engine to use full rank states. A pair of qubits (labelled \textbf{A} and \textbf{B}) are taken as the working medium (system) whose evolution is governed by a Hamiltonian  $H=H_{loc}+H_{int}$ given in Eq.(\ref{hloc}) and Eq.(\ref{hint}).  
\begin{figure*}[htp!]
    \centering
    \includegraphics[width=2\columnwidth]{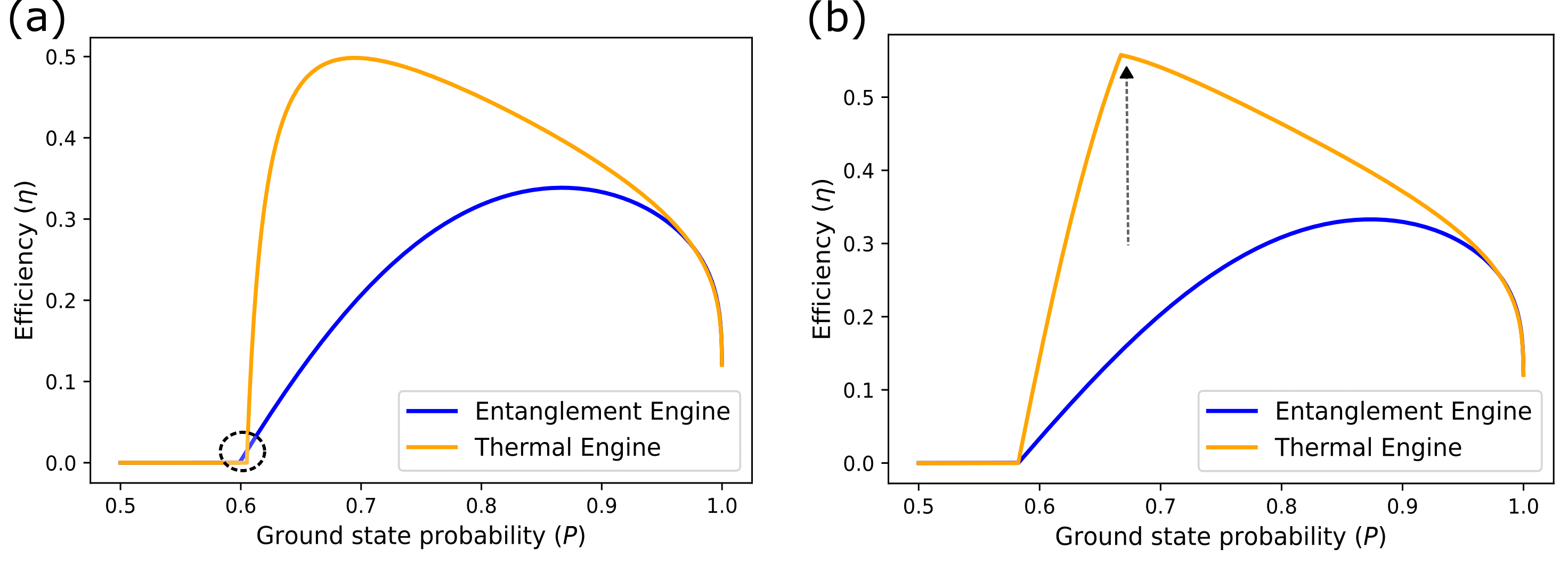}
    \caption{Efficiency as a function of pointer ground state probability~$P$, for maximally-faithful and (a) non-invasive measurement, (b) unbiased. Parameters of the simulation are $\omega_A = 10$, $\delta = 50$, $\theta = \pi/5$, $\beta E_p=1/30$, and $q=0.05$. The crossover of efficiency in (a) highlighted with the dotted circle, and the kink in panel~(b) indicated by the dotted arrow are both discussed in the main text.}
    \label{fig:1}
\end{figure*}
To extract work from system we consider a positive detuning $\delta=\omega_B-\omega_A$. The initial state of the system is taken as
\begin{align}
\rho_{\mathrm{2qb}}(0) = (1-q) \ketbra{10} + q \dfrac{\mathrm{e}^{-\beta \mathrm{H_{loc}}}}{{Z_\mathrm{loc}}}
, \quad 0<q\leq 1,
\label{rhoqbtwo}
\end{align} 
where $Z_\mathrm{loc} = \Tr[exp\{-\beta H_{loc}\}]$. The state $\rho_{2qb}(0)$ is a convex mixture of the pure state and thermal state. The admixture probability $q>0$ in Eq.(\ref{rhoqbtwo}) guarantees the state to be full rank, circumventing the aforementioned infinite cooling cost. We let the system evolve from the initial state $\rho_\mathrm{2qb}(0)$ for time $t_1 = \pi/\sqrt{g^2+\delta^2}$, such that the system state becomes maximally entangled (given the constraints on its eigenspectrum)
\begin{align}
\rho_\mathrm{2qb}(t_1) &= (1-q) \{\cos^2\theta \ketbra{10} + \sin^2\theta\ketbra{01} \nonumber \\ &- \sin\theta\cos\theta (\ketbra{10}{01} + \ketbra{01}{10})\} \nonumber \\ &+ q  \dfrac{\mathrm{e}^{-\beta \mathrm{\tilde{H}_{loc}}}}{{\tilde{Z}_\mathrm{loc}}}
\label{rho2qbt}.
\end{align}
Here $\tilde{H}_{loc}= U_{l}H_{loc}U^\dag_l$,  $U_l=\exp({-i H_{2qb}t_1})$,\\  $\tan\theta=g/\delta$ and $\tilde{Z}_{loc}= \Tr[exp\{-\beta \tilde{H}_{loc}\}]$. At this point we switch off the interaction potential which costs energy \begin{equation}
   E_{prep}=-\Tr[\rho_{2qb}(t_1) H_{int}]. 
\end{equation}
As we are interested in single excitation subspace, we can use a single qubit pointer in thermal state $\tau(\beta')$, to measure qubit \textbf{B}. The pointer is governed by Hamiltonian $H_P=E_P\ketbra{1}{1}$. 
To make the measurement non-invasive (or unbiased) we use $U_{corr}=I\otimes U_{CNOT}$ (or $U_{corr}=I\otimes U_{UNB}$, where $U_{UNB}=|00\rangle\langle00|+| 01\rangle\langle 11|+| 11\rangle\langle 10|+| 10\rangle\langle01|$) on the system-pointer state to correlate pointer with qubit \textbf{B}. This process consumes energy given by Eq.(\ref{ecor}). The work is extracted by applying a SWAP operation on the system qubits ($U_{swap}=\ketbra{00}+\ketbra{01}{10}+\ketbra{10}{01}+\ketbra{11}$) if the pointer is observed in excited state. The system is then reset to $\rho_{2qb}(0)$, to repeat the cycle again. In order to make the measurements more accurate the pointer needs to be cooled from $\tau(\beta)$ to $\tau(\beta')$, before correlating it with the system. The free energy difference associated with cooling can be expressed as~\cite{clivaz2019unifying,taranto2021landauer} 
\begin{equation}
E_{cool}=\left(\frac{\beta'}{\beta}-1
\right)\left(\frac{E_P}{1+e^{-\beta' E_P}}-\frac{E_P}{1+e^{-\beta  E_P}}\right)
\label{cool}.
\end{equation} 

\noindent To contrast this against thermal correlations, we now study a thermodynamically-consistent TCME and demonstrate that such an engine can be designed to have better efficiency while having same work output as the entanglement engine. Consider the initial state 
\begin{align}
\Tilde{\rho}_\mathrm{2qb} &= (1-q) \{\cos^2\theta \ketbra{10} + \sin^2\theta\ketbra{01}\} \nonumber \\  &+ q  \dfrac{\mathrm{e}^{-\beta \mathrm{H_{loc}}}}{{Z_\mathrm{loc}}}
\label{rho2qbe},
\end{align}
where $\tan^2\theta=n_{th}/(n_{th}+1)$ and $n_{th}$ is average population defined below. The process of work extraction from the TCME is same as that of the EME, but a difference arises in the reset step of both engines.
The EME is reset to a reference state Eq.\eqref{rhoqbtwo} and the TCME is reset arbitrarily close to Eq.\eqref{rho2qbe} in finite time dissipatively~\cite{breuer2002theory}
\begin{align}
\frac{d\rho}{dt}&=-i[H_{loc},\rho]+\gamma D\left[\sigma^+_A|0\rangle\langle0|_B\right]\rho+\gamma D\left[|1\rangle\langle1|_A\sigma^-_B\right]\rho\nonumber\\&+\gamma'n_{th}D\left[|01\rangle\langle10|\right]\rho+\gamma' (n_{th}+1) D\left[|10\rangle\langle01|\right]\rho.\label{rhodot}
\end{align}
Here $\mathcal{D}[L]X \equiv L X L^{\dagger}-\frac{1}{2}\left\{L^{\dagger} L,X\right\}$ is the usual Lindblad dissipator, $\{\gamma,\gamma'\}$ are transition rates and $n_{th}=1/(e^{\beta(E_B-E_A)}-1)$. We choose a different initial state for the TCME as the state can be prepared and maintained using dissipators given in Eq.(\ref{rhodot}). 
This consumes less energy in the preparation and reset step than EME. The energy spent in resetting (preparing) classical correlations will consume an amount of energy given by Eq.\eqref{ereset}. 

The work extracted by starting from state Eq.\eqref{rho2qbt} and Eq.\eqref{rho2qbe} differs by an amount proportional to the thermal admixture $q$. For $q\ll1$ and $\beta>1$, the difference in the work extracted for TCME and EME can be made arbitrarily small to compare the efficiency of both engines directly.
We plot the efficiency versus purity of pointer for two different types of measurement in Fig.~\ref{fig:1}. Efficiency is evaluated as $\eta=-W/E_{meas}$, where $E_{meas}$ is the total energy used in measurement, i.e., $E_{meas}=E_{prep}+E_{cool}+E_{corr}+E_{reset}$. We begin by noting that the graphs do not start at the origin because work output in such settings will be negative. It can be seen that the engine with the thermal initial state outperforms the engine with the entangled initial state everywhere, except for a small region near $p=0.6$ for non-invasive measurement.
This is due to the fact that the energy required in resetting the thermal state is larger than the energy required for resetting an entangled initial state for low values of purity. This region of enhanced performance by the engine with initially entangled state decreases with decreasing the admixture probability $q$ and vanishes in the limit of $q\to 0$. As the energy costs $E_{corr}$ and $E_{cool}$ are similar in both engines, the observed difference in efficiency arises due to the difference in energy required to reset the states and to prepare the entangled initial state.
We also note the sharp kink in the performance of the thermal engine with unbiased measurement. This is because the reset cost is negative before the kink and positive for larger $P$. The negative value indicates that energy is released while resetting the system. This cost is only accounted in the efficiency if the value is positive and considered as zero if negative. If the reset cost is negative, one can choose to extract the energy as work, thereby enhancing the efficiency of the TCME. However, even without extracting this additional energy as work, the TCME has better efficiency than EME as shown in Fig.~\ref{fig:1}.


\section{\label{sec:V}Discussion} 


\noindent Past approaches towards measurement engines have assumed idealized measurements \cite{bresque2021two,PhysRevA.104.062210,rossnagel2014nanoscale,PhysRevResearch.4.L012029}. As such measurements are thermodynamically infeasible, one needs to model measurement with full rank pointer states. In this manuscript, we considered a more accurate measurement model involving states of maximal rank for both pointer and primary system. We compared the performance of different initial states, one with entanglement and another one with thermal correlations. The realistic measurement model results in more work extraction with increasing pointer purity, but the associated cost of increasing purity eventually decreases the efficiency of the engine, which sets a trade-off between extracted work and efficiency. We see that the performance of the engine using entangled states can be matched using only classical correlations. We observe that the required purity of the pointer to operate the engine most efficiently is less in the case of thermal correlations than for entangled correlations. Hence a lower temperature bath is needed for optimal performance of the entangled engine than for the corresponding thermally-correlated engine.
We note that our results are consequences of the properties of measurement and hence valid for any general measurement engine where work is extracted by applying feedback (i.e., based on information obtained from measurement). 

\begin{acknowledgments}
 FCB acknowledges funding by the Irish Research Council under grant number IRCLA/2022/3922 and the John Templeton Foundation (grant no. grant 62423). SV acknowledges support from a DST-SERB Early Career Research Award (ECR/2018/000957) and DST-QUEST grant number DST/ICPS/QuST/Theme-4/2019. SV thanks A. Werulkar for useful discussions. 
\end{acknowledgments}

\end{document}